\documentclass[preprint]{ptephy_v1}

\preprintnumber{OU-HET-863} 

\usepackage{amsmath} 
\usepackage{amsthm} 
\usepackage{graphics} 
\usepackage{braket}

\newcommand{\Slash}[1]{{\ooalign{\hfil/\hfil\crcr$#1$}}}
\numberwithin{equation}{section}
\def\Rnum#1{\uppercase\expandafter{\romannumeral #1}} 

\begin{document}

\title{Renormalization of Extended QCD$_2$}


\author{Hidenori Fukaya}
\author{Ryo Yamamura}
\affil{Department of Physics, Osaka University, Toyonaka, Osaka 560-0043 Japan
\email{hfukaya@het.phys.sci.osaka-u.ac.jp, ryamamura@het.phys.sci.osaka-u.ac.jp}
}




\begin{abstract}%
Extended QCD (XQCD) proposed by Kaplan \cite{Kaplan:2013dca} 
is an interesting reformulation of QCD with additional bosonic auxiliary fields. 
While its partition function is kept
exactly the same as that of original QCD,
XQCD naturally contains properties of low energy hadronic models.
We analyze the renormalization group flow of two-dimensional (X)QCD,
which is solvable in the limit of large number of colors $N_c$,
to understand what kind of roles the auxiliary degrees of freedom play
and how the hadronic picture emerges
in the low energy region.
\end{abstract}

\subjectindex{B00, B06, B32, B34, B35}

\maketitle

\section{Introduction}
\label{sec:Intro}
In Ref.~\cite{Kaplan:2013dca}, Kaplan proposed an interesting reformulation of QCD named as Extended QCD or XQCD.
This new formulation contains additional auxiliary bosonic fields, 
keeping the partition function of QCD unchanged.
The physics of XQCD is exactly the same as that of QCD, 
as long as the source operators of the ordinary quark and gluon fields are inserted.

It is shown in Ref.~\cite{Kaplan:2013dca} that XQCD can describe several low energy hadronic pictures
more naturally than QCD itself (in the limit of large number of colors $N_c$, 
where it is particularly simple to understand). 
The remarkable difference comes from the vacuum expectation value (VEV) of the auxiliary scalar field.
This VEV directly gives the constituent mass to the quarks, 
which is an essential part of the quark models, 
and at the same time, makes the pseudo-scalar propagator massless,
whose non-linear chiral transformation has exactly the same representation 
as the one in chiral perturbation theory.
Moreover, it can be explained how the VEV is weakened by the presence of the baryonic source, 
the property suggested by the bag models~\cite{Chodos:1975ix}.


The purpose of this paper is to understand what kind of roles auxiliary degrees of freedom play
in the low energy region more concretely.
It would be interesting if one could simulate lattice XQCD
in four-dimensions and directly examine the above features.
Unfortunately, the current formulation of XQCD suffers
from the sign problem even with zero chemical potential.
Instead, we study the two-dimensional version of (X)QCD
(we will simply denote QCD$_2$ or XQCD$_2$ in the following),
in the large $N_c$ limit.
This theory 
is known as the 't Hooft model~\cite{'tHooft:1974hx}
whose exact solution for quark propagator (in a particular gauge) 
and numerical solutions for meson masses given non-perturbatively.
The advantage of studying the 't Hooft model is that the theory is 
particularly simplified in the large $N_c$ limit and solvable. 
We consider this work as the first step to future studies of four-dimensional (X)QCD with $N_c=3$.

In this work, we study the Wilsonian renormalization group (RG) flow of XQCD$_2$.
We find that the auxiliary fields become dynamical
when we take into account quantum corrections.
Note that
the degrees of freedom in XQCD should be the same as those in QCD,  
since auxiliary fields can give no effects on the original theory.
Thus we can interpret the  ``dynamical auxiliary field" as 
just a transmutation of the degrees of freedom in QCD. 
In particular,
the (pseudo)scalar auxiliary field
should play a key role in the low energy effective action.
It contains the degrees of freedom of pions,
the lightest hadrons, as a consequence of the dynamical chiral symmetry breaking~\cite{Nambu:1961tp}.

We also find that XQCD provides an interesting extension 
of the renormalization ``scheme''.
When we compute the RG flow, 
we usually restrict ourselves to the space of the original fields given in our Lagrangian.
In the case of QCD, for example, 
we only consider running of the couplings among quarks and gluons.
However, in XQCD, we can insert at an arbitrary scale $\Lambda_{\rm cut}$
new bosonic degrees of freedom and the RG flow is
extended to the space of their new interactions.
Note that a similar idea was already tried in the works on the ``dynamical hadoronization''~\cite{Braun:2014ata}. 
They converted the four-quark interactions, which were
developed along the conventional RG flow of QCD,
into the mesonic fields.
But XQCD has a wider possibility in that no source of the original 
(four-quark) interaction is required. 
The scale(s)  $\Lambda_{\rm cut}$('s) and 
the number of mesonic degrees of freedom are completely arbitrary.
It is also important to note that XQCD has no risk of overcounting
the physical degrees of freedom in original QCD.

This highly extended ``scheme'' of renormalization suggests
many interesting applications beyond QCD.
Since the number of auxiliary fields and
its scale $\Lambda_{\rm cut}$ are arbitrary, 
``one'' theory has infinitely many different effective actions 
at low energy, which are all physically equivalent.
Moreover, the scheme suggests that there could exist 
a cut-off $\Lambda_{\rm cut}$ of the effective theory, 
which has no physical meaning.
These aspects may give new insights to the current problems
of the particle theory, such as naturalness problem.
We would like to discuss these new possibilities in detail.


The rest of our paper is organized as follows.
First, we review the formulation of XQCD including its two-dimensional version,
and how it shows low energy hadronic pictures
in the large $N_c$ limit in Sec.~\ref{sec:reviewXQCD}. 
Then we explain our renormalization ``scheme'' in Sec.~\ref{sec:RGQCD2}.
Finally we compare the RG flow 
of QCD$_2$ and XQCD$_2$ in the large $N_c$ limit.
A summary is given in Sec.~\ref{sec:conclusion}.

\section{Extended QCD and its two-dimensional version}
\label{sec:reviewXQCD}
In this section, we review the original Extended QCD \cite{Kaplan:2013dca} 
in four dimensions and construct its two-dimensional version.
We also summarize what is known in this two-dimensional large $N_c$ QCD (the 't Hooft model).

\subsection{XQCD in four dimensions}
\label{subsec:defXQCD}

We consider QCD with $N_f$ flavors of quarks and gauge group $SU(N_c)$ 
in four-dimensional Euclidean spacetime :
\begin{align}
S_{\text{QCD}}
=&~
N_c\int d^4x~\left[\bar{\psi}_{ia}(\Slash{D}+m)^{a}_{~b}\psi^{ib} 
+\frac{1}{4g^2}\text{Tr}~\mathbf{F}_{\mu\nu}\mathbf{F}_{\mu\nu}\right],
\label{eq:QCD}
\end{align}
where $\Slash{D}=\gamma^\mu(\partial_\mu+i\mathbf{A}_\mu)$ is the covariant derivative, 
and $\mathbf{A}_\mu$ denotes the gluon field.
Here $a,b,\dots$ are color indices and $i,j,\dots$ are flavor indices.




XQCD is defined by introducing three types of auxiliary fields,
the scalar field $\Phi$, vector $\mathbf{v}_\mu$ and axial vector $\mathbf{a}_\mu$,
with the action in a Gaussian form,
\begin{align}
S_{\text{aux}}[\Phi,\Phi^\dag,\mathbf{v}_\mu,\mathbf{a}_\mu]
=
&N_c\lambda^2\int d^4x 
\bigg[\frac{}{}
\text{Tr}~(\Phi^\dag+2\lambda^{-2}\bar{\psi}_aP_+\psi^a)
(\Phi+2\lambda^{-2}\bar{\psi}_aP_-\psi^a) \notag \\
&~~~~~~~~~~~~~
+\frac{1}{2}\text{Tr}~({\mathbf v}_\mu+\lambda^{-2}\bar{\psi}_i\gamma_\mu\psi^i)
({\mathbf v}_\mu+\lambda^{-2}\bar{\psi}_i\gamma_\mu\psi^i) \notag \\
\label{eq:gaussian}
&~~~~~~~~~~~~~
+\frac{1}{2}\text{Tr}~({\mathbf a}_\mu+i\lambda^{-2}\bar{\psi}_i\gamma_\mu\gamma_5\psi^i)
({\mathbf a}_\mu+i\lambda^{-2}\bar{\psi}_i\gamma_\mu\gamma_5\psi^i)
\bigg],
\end{align}
which keeps the original QCD partition function intact (up to a constant) :
\begin{align}
Z_{\text{QCD}}
=
&~
\int D\psi D\bar{\psi} D\mathbf{A}_\mu~e^{-S_{\text{QCD}}[\psi,\bar{\psi},\mathbf{A}_\mu]} \notag \\
=
&~
\int D\psi D\bar{\psi} D\mathbf{A}_\mu
D\Phi D\Phi^\dag D\mathbf{v}_\mu D\mathbf{a}_\mu
~e^{-S_{\text{QCD}}[\psi,\bar{\psi},\mathbf{A}_\mu]
-S_{\text{aux}}[\Phi,\Phi^\dag,\mathbf{v}_\mu,\mathbf{a}_\mu]} \notag \\
\equiv
&~
Z_{\text{XQCD}}.
\end{align}
Here, the color singlet $\Phi$ transforms as a bifundamental representation 
under the $SU(N_f)_L\times SU(N_f)_R$ chiral symmetry,
and the flavor singlet $\mathbf{v}_\mu$ and $\mathbf{a}_\mu$ 
are $N_c\times N_c$ matrices (the singlet plus
adjoint representations of the $SU(N_c)$ gauge group).

Note that each term of the action Eq.~(\ref{eq:gaussian})
has a non-renormalizable four-quark interaction.
However, they automatically cancel through the Fierz identity
\begin{equation}
\label{eq:fierzeu}
(P_+)_{mn}(P_-)_{m'n'}+(P_-)_{mn}(P_+)_{m'n'}=
\frac{1}{4}[(\gamma_\mu)_{mn'}(\gamma_\mu)_{m'n}
-(\gamma_\mu\gamma_5)_{mn'}(\gamma_\mu\gamma_5)_{m'n}],
\end{equation}
where $P_{\pm}=\frac{1}{2}(1\pm\gamma_5)$.
Therefore, our new theory
\footnote{
In general, 
scalar field $\Phi$ is a complex matrix. 
For $N_f=2$, 
since the fundamental representation of $SU(2)$
is a pseudo-real representation,
we can impose the reality condition 
$\Phi=\sigma_2\Phi^\ast\sigma_2$
to $\Phi$.
In this case the factor $\frac{1}{2}$ is needed in front of the 
mass term of $\Phi$.
} :
\begin{align}
S_{\text{XQCD}}
=
N_c\int d^4x~\bigg[\bar{\psi}(\mathcal{D}+m)\psi 
&~+\frac{1}{4g^2}\text{Tr}~\mathbf{F}_{\mu\nu}\mathbf{F}_{\mu\nu} \notag \\
\label{eq:XQCD}
&~+\lambda^2\left(\text{Tr}~\Phi^\dag\Phi
+\frac{1}{2}\text{Tr}~[{\mathbf v}_\mu{\mathbf v}_\mu
+{\mathbf a}_\mu{\mathbf a}_\mu]\right)\bigg],
\end{align}
where 
\begin{equation}
\label{eq:diracopeXQCD}
\mathcal{D}\equiv
\Slash{D}+\Slash{{\mathbf v}}+i\Slash{{\mathbf a}}\gamma_5+
2(\Phi P_++\Phi^\dag P_-),
\end{equation}
is manifestly renormalizable.

Here, $\lambda$ is an arbitrary parameter which has a mass dimension.
Also, we can define the bare XQCD action at an arbitrary scale
$\Lambda_{\rm cut}$.
Therefore, we have introduced two unphysical scales.
Of course, any physical observables cannot depend on
$\lambda$ nor $\Lambda_{\rm cut}$.
As explained below, the natural choice for the former
value is the QCD scale, $\lambda\sim \Lambda_{\rm QCD}$,
while we want $\Lambda_{\rm cut}$ to be at higher energy 
near the real cut-off of the theory.
Later we will discuss that this high ambiguity introduced
in XQCD gives the extension of the ``scheme'' of the renormalization.

Since the integration over auxiliary fields is just a constant, 
the expectation value of any operator 
involving gluon and quark fields only, 
is equivalent to that of QCD :
\begin{equation}
\label{eq:VEVsame}
\braket{\mathcal{O}(\psi,\bar{\psi},\mathbf{A}_\mu)}_{\text{XQCD}}
=
\braket{\mathcal{O}(\psi,\bar{\psi},\mathbf{A}_\mu)}_{\text{QCD}}.
\end{equation}
This makes a big contrast to the previous attempts of
simply adding scalar fields to QCD \cite{Brower:1994sw,Brower:1995vf,Kogut:2004ia}.
Since they are formally different from QCD, and
have non-renormalizable four-quark interactions,
it is non-trivial to keep the theory in the same universality 
class of QCD.
In this respect, XQCD, which is exactly equivalent to QCD, 
has a theoretically firmer background.

Although XQCD and QCD are equivalent, 
their Feynman diagrams are quite different.
The striking difference is seen when we 
assume a non-zero VEV to the chiral condensate.
Since $\Phi$ shares the same quantum numbers as the scalar quark bilinear operator,
it should also have VEV.
In Ref.~\cite{Kaplan:2013dca}, it is explicitly computed in the large $N_c$ limit as
\begin{equation}
\label{eq:VEV}
\lambda^2\braket{\Phi^i_{~j}}_{\text{XQCD}}
=-\delta^i_{~j}\braket{\bar{\psi}\psi}_{\text{QCD}}
\equiv \delta^i_{~j}\Sigma .
\end{equation}
This VEV directly gives the constituent mass $M=2\Sigma/\lambda^2$ to the quarks. 

It is also important to note the relative $i$ 
between the $\mathbf{v}_{\mu}$ and $\mathbf{A}_{\mu}$ couplings 
in Eq.~\eqref{eq:diracopeXQCD}.
It means that 
the exchange of $\mathbf{v}_{\mu}$ is repulsive  
while that of gluon is attractive.
When the exchange of $\mathbf{a}_{\mu}$ is also taken into account,
the repulsion is specifically between right-handed and left-handed quarks.
Hence, the exchanges of vector and axial vector auxiliary fields
(partially) weaken the attractive gluon exchanges.
The introduction of the scalar auxiliary field $\Phi$, 
which gives the constituent mass to the quarks, 
is concomitant with weakening of the interaction between quarks. 
This property is what assumed in the quark model~\cite{GellMann:1964nj,Zweig:1981pd},  
described by weakly interacting massive quarks. 
In this way, XQCD naturally contains the feature of the quark model, 
which can not be explained by original QCD.
As the $\rho$ meson is made by two constituent quarks,
an optimal choice \cite{Kaplan:2013dca} of $\lambda$ is around $300$ MeV.

Moreover, it is shown in  \cite{Kaplan:2013dca}  that
the above quark model picture is compatible with 
the presence of the light pions as the 
(pseudo) Nambu-Goldstone (NG) bosons.
Having the heavier constituent mass,
the quark's connected diagrams cannot have a long-range 
correlation. Instead, XQCD explicitly includes the propagation
of $\Phi$ containing the pionic mode in it.
Thus, XQCD diagrammatically distinguishes the pions from other
mesons made by constituent quarks.




\subsection{Application to the 't Hooft model}
\label{subsec:tHooft}


In this work, we consider the large $N_c$ limit of QCD in two-dimensional Lorentzian spacetime,
which is the so-called 't Hooft model~\cite{'tHooft:1974hx}.
An exact solution for the quark propagator 
and numerical solutions for the meson masses are known.
Solvability of the theory comes from the fact that 
gauge fields have only two degrees of freedom in two dimensions
and we can eliminate the self-interaction of gauge fields by a suitable gauge fixing. 
The elimination of the self-interaction dramatically simplifies the theory in the large $N_c$ limit. 
In addition to the simplicity,
two-dimensional gauge theories show the confinement and 
the chiral symmetry breaking in the large $N_c$ limit.
Thus, the 't Hooft model is a good test ground for QCD. 
In this subsection, we briefly review this model and construct its extended version.

Let us first introduce
the light-cone coordinate 
\begin{equation}
\label{eq:LCcoor}
x^{\pm}=(x^0\pm x^1)/\sqrt{2}.
\end{equation}
With this, the metric is given by
\begin{equation}
\label{eq:LCmetric}
g^{+-}=g^{-+}=g_{+-}=g_{-+}=1,
\end{equation}
and all other components are zero.
Note that $x^2=x^\mu x_\mu=2x^+x^-$.

Next, we take the light-cone gauge :
\begin{equation}
\label{eq:gaugecond}
\mathbf{A}_-=\mathbf{A}^+=0.
\end{equation}
This choice of gauge is Lorentz invariant,
since its transformation 
is operated multiplicatively on each coordinate.
With this gauge, the QCD Lagrangian is given in a simple form
\begin{align}
\mathcal{L}
\label{eq:QCDlagLC}
=
\frac{1}{2}\text{Tr}~(\partial_-\mathbf{A}_+)^2
+\bar{\psi}i\Slash{\partial}\psi-m\bar{\psi}\psi
-\frac{g}{\sqrt{N_c}}\bar{\psi}\mathbf{A}_{+}\gamma^+\psi.
\end{align}
Note that there is no self-interaction term among gluons.

Here $\gamma^\pm$ are the gamma matrices 
satisfying
\begin{align}
\label{eq:gammaLC}
(\gamma^+)^2=(\gamma^-)^2=0~,~~
\{\gamma^+,\gamma^-\}=2.
\end{align}
It is also useful to define
\begin{equation}
\gamma_3
=
-\gamma^0\gamma^1
=
\frac{1}{2}[\gamma^+,\gamma^-],
\end{equation}
which is the counterpart of $\gamma_5$ in four dimensions.

The Feynman rule is given by the gluon propagator, 
the vertex factor of the quark-antiquark-gluon interaction and the quark propagator,
\begin{equation}
D_{\mu\nu}(k)
=
i\delta_{\mu+}\delta_{\nu+}\frac{1}{(k_-)^2},
\end{equation}
\begin{equation}
-\frac{ig\gamma^+}{\sqrt{N_c}},
\end{equation}
\begin{equation}
\mathbf{S}_{\text{tree}}(p)
=
i~\frac{p_+\gamma^++p_-\gamma^-+m}
{2p_+p_--m^2+i\epsilon}.
\end{equation}
Since every gluon-quark vertex contains $\gamma^+$,
the internal quark line is always 
sandwiched by two $\gamma^+$s. 
Since
\begin{align}
\label{eq:gammaLCsand}
\gamma^+\left\{
\begin{array}{c}
1 \\
\gamma^+\\
\gamma^- \\
\gamma_3
\end{array}
\right\}\gamma^+
=
2\gamma^+\left\{
\begin{array}{c}
0 \\
0 \\
1 \\
0
\end{array}
\right\},
\end{align}
we only need to consider $\gamma^-$ component of the quark propagator.

\begin{figure}[htbp]
\begin{center}
\includegraphics[width=7cm]{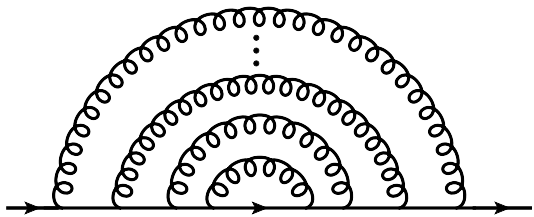}
\caption{A rainbow diagram contributing to quark propagator.}
\label{fig:rainbow}
\end{center}
\end{figure}
\begin{figure}[htbp]
\begin{center}
\includegraphics[width=7cm]{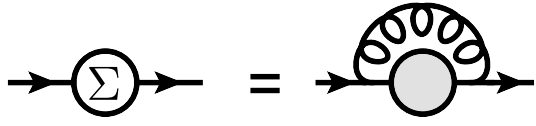}
\caption{A diagrammatic expression of the self-consistent equation for the self-energy $\Sigma$(p).}
\label{fig:selfenergy}
\end{center}
\end{figure}

Thanks to this simple Feynman rule and the large $N_c$ limit,  
we can non-perturbatively compute the quark self-energy.
The quantum corrections to the quark propagator in the large $N_c$ limit
are expressed by the so-called rainbow diagrams shown in Fig.~\ref{fig:rainbow},
which is proportional to $\gamma^+$.
\vspace{1em}
Now the ``full'' quark propagator is expressed as
\begin{align}
\label{eq:fullpropn}
S(p) = \frac{ip_-}
{2p_+p_--m^2-p_-\Sigma(p)+i\epsilon},
\end{align}
and we obtain a self-consistent equation (see Fig.~\ref{fig:selfenergy})
\begin{align}
\label{eq:SCeq}
-i\Sigma(p)
=
-4ig^2\int
\frac{dk_+dk_-}{(2\pi)^2}
S(p-k)\frac{1}{(k_-)^2}. 
\end{align}
Note that the integration has the IR divergence at $k_-\rightarrow 0$.
According to Ref~\cite{col}, 
let us take the principle-value prescription and
obtain 
\begin{equation}
\label{eq:exactsol}
p_- \Sigma(p)=-\frac{g^2}{\pi}.
\end{equation}

This result may look pathological since the constituent quark mass squared
\begin{equation}
M^2 = m^2-g^2/\pi,
\end{equation}
becomes tachyonic when $g$ is strong.
It is, however, regarded as just an artifact of the gauge fixing and the IR regularization.
In fact, the other choice of the IR regularization, which give a positive values of $M^2$,
does not change the meson spectrum~\cite{Bars:1977ud,Li:1987hx}.\\

To ``extend'' the 't Hooft model is
almost straightforward as the original XQCD in four-dimensions.
However,
there are two different points to be minded.
One is the difference of the Fierz identity,
which depends on dimensions.
Another is the signature of the spacetime metric:
to employ the light-cone gauge,
we have to work in a Lorentzian spacetime,
although original XQCD in Sec.~\ref{sec:reviewXQCD} is defined 
in four-dimensional Euclidean spacetime.

The Fierz identity of two-dimensional theories is 
\begin{align}
\label{eq:fierz2d}
(P_+)_{mn}(P_-)_{m'n'}
+(P_-)_{mn}(P_+)_{m'n'}
=
\frac{1}{2}(\gamma_\mu)_{mn'}(\gamma^\mu)_{m'n},
\end{align}
where $\gamma^0$ and $\gamma^1$ are taken to be hermitian and anti-hermitian respectively.
Projection matrices $P_{\pm}$ are defined by $P_{\pm}=(1\pm\gamma_3)/2$.
Note that there is no axial vector in two-dimensions.
We can write the identity with quark fields such that
\begin{align}
\label{eq:2dfierzquark}
\left(
\bar{\psi}_{ja}P_+\psi^{ia}
\right)
\left(
\bar{\psi}_{ib}P_-\psi^{jb}
\right)
=
-\frac{1}{4}
\left(
\bar{\psi}_{ib}\gamma_\mu\psi^{ia}
\right)
\left(
\bar{\psi}_{ja}\gamma^\mu\psi^{jb}
\right).
\end{align}

Next, let us consider the auxiliary field path integral in the Lorentzian space-time,
\begin{align}
\int\mathcal{D}\phi~ e^{iS(\phi)}.
\end{align}
Unlike the Euclidean case, it is not necessary for $S(\phi)$ to be positive
because of the existence of the factor $i$.
This means that there is some ambiguity 
in introducing the auxiliary fields.
In this work,
we require a condition that the
mass terms of the scalar and spacial part of the vector auxiliary fields 
are not tachyonic, at least, at the tree level, and then obtain
\begin{align}
\label{eq:auxiPI}
e^{iS_{\rm aux}[\Phi,\Phi^\dagger, {\mathbf v}_\mu]} =
&\exp
\left[-i\lambda^2\int d^2x 
\left\{\frac{}{}
\text{Tr}~\left(\Phi^\dag+\frac{\sqrt{2}}{\sqrt{N_c}}\frac{\alpha}{\lambda}\bar{\psi}_aP_+\psi^a\right)
\left(\Phi+\frac{\sqrt{2}}{\sqrt{N_c}}\frac{\alpha}{\lambda}\bar{\psi}_aP_-\psi^a\right)\right. \right. 
\notag \\
&~~~~~~~~~~~~~\left.\left.
-\frac{1}{2}\text{Tr}~\left({\mathbf v}_\mu+\frac{1}{\sqrt{N_c}}\frac{\alpha}{\lambda}
\bar{\psi}_ii\gamma_\mu\psi^i\right)
\left({\mathbf v}^\mu+\frac{1}{\sqrt{N_c}}\frac{\alpha}{\lambda}
\bar{\psi}_ii\gamma^\mu\psi^i\right)
\right\}
\right]\nonumber\\
=&\exp\left[
i\int d^2x~\left\{
-\frac{\alpha\lambda}{\sqrt{N_c}}
\bar{\psi}[\sqrt{2}(\Phi P_++\Phi^\dag P_-)
-
i\Slash{\mathbf{v}}
]\psi
-\lambda^2\left(\text{Tr}~\Phi^\dag\Phi
-\frac{1}{2}\text{Tr}~{\mathbf v}_\mu{\mathbf v}^\mu
\right)\right\}\right],
\end{align}
where $\lambda$ and $\alpha$ are arbitrary real parameters.
The mass dimensions of auxiliary fields and parameters are given by
\begin{align}
[\Phi]=[\mathbf{v}_{\mu}]=0~,~~[\alpha]=0~,~~[\lambda]=1.
\end{align}
The total action of XQCD$_2$ is given by
\begin{align}
S_{\text{XQCD}}
=
\int d^2x~\bigg[&~\bar{\psi}[\mathcal{D}'-m]\psi 
+\frac{1}{2}\text{Tr}~(\partial_-\mathbf{A}_+)^2 
\label{eq:XQCD2shift}
-\lambda^2\bigg(\text{Tr}~\Phi^\dag\Phi
-\frac{1}{2}\text{Tr}~{\mathbf v}_\mu{\mathbf v}^\mu\bigg)
\bigg],
\end{align}
where
\begin{equation}
\label{eq:diracopeXQCD2}
\mathcal{D}'\equiv
i\Slash{\partial}
-\frac{g}{\sqrt{N_c}}\mathbf{A}_+\gamma^+
+\frac{i\alpha\lambda}{\sqrt{N_c}}\Slash{{\mathbf v}}
-\frac{\sqrt{2}\alpha\lambda}{\sqrt{N_c}}
(\Phi P_++\Phi^\dag P_-).
\end{equation}

In the above action, the mass term of quarks 
is the only source of explicit breaking of the chiral symmetry.
We can absorb this symmetry breaking in the $\Phi$'s shift:
\begin{align}
\label{eq:shift}
\Phi
\rightarrow
\Phi-\frac{\sqrt{N_c}}{\sqrt{2}\alpha\lambda}m.
\end{align}
Then the fermion mass term is converted to
\begin{equation}
\label{eq:massconvert}
-m\bar{\psi}\psi \to \frac{\sqrt{N_c}}{\sqrt{2}}\frac{\lambda}{\alpha}m\text{Tr}~(\Phi+\Phi^\dag).
\end{equation}
We also use this re-definition of the mass term in the RG studies of XQCD$_2$. 



\section{Extended renormalization scheme}
\label{sec:RGQCD2}

As explained above, although QCD and XQCD are exactly equivalent,
their low energy expressions are expected to be different.
To understand this more clearly, we perform the Wilsonian 
renormalization group transformation on both theories and 
compare their low energy effective actions.

We would like to address two possible features of XQCD.
One is how the mesonic degrees of freedom become dynamical.
As $\Phi$ is expected to play a role of the NG boson at low energy,
the RG flow should develop its kinetic term at low energy,
keeping its mass near zero.
Another issue is to see what happens on the original quark 
and gluon sectors along the RG flow.
As hadrons play more important roles at low energy,
the original quarks and gluons should decrease
their relevance,
and can eventually be decoupled from the effective action,
near the scale of their (constituent) masses.
We may be able to see this as cancellation with 
the vector auxiliary fields.

The inclusion of the auxiliary fields extends the (relevant)
parameter space of the theory.
The new terms of the effective Lagrangian we should consider are
\begin{eqnarray}
{\rm Tr}\partial_\mu \Phi^\dagger \partial^\mu \Phi,\;\;
{\rm Tr}\partial_\nu {\bf v_\mu} \partial^\nu {\bf v}^\mu,\;\;
{\rm Tr}(\partial_\mu  {\bf v^\mu})^2, \;\;
\text{Tr}~\Phi^\dag\Phi,\;\;
\text{Tr}~{\mathbf v}_\mu{\mathbf v}^\mu,\;\;
\bar{\psi}(\Phi P_++\Phi^\dag P_-)\psi,\cdots
\end{eqnarray}
However, as the original theory has only two parameters $g$ and $m$,
the new interactions are not independent,
but essentially controlled by these two parameters.
Namely, the RG flows are restricted 
on a two-dimensional surface in the extended parameter space.

Which two-dimensional surface we take is determined by
the choice of the regularization we use, 
and the re-definition of the coupling constants 
(by giving counterterms).
In the view of RG flow of $N$ parameters,
$N-2$ constraints can be given by these counterterms.
Therefore, the choice of the surface corresponds to 
nothing but the choice of the renormalization scheme.
Thus, XQCD can be regarded as the extension of the renormalization
scheme to the extended theory space\footnote{
Note that extending the theory space and giving constraints on it, 
are widely used (sometimes unconsciously)
even in the conventional RG analyses.
For example, when we compute the renormalization of a supersymmetric theory,
we have to employ some regularization which breaks the symmetry,
and natural RG flows go through the non-supersymmetric space.
We could still expect non-trivial cancellations of the 
contributions in that space, so that the theory remains to be supersymmetric.
However, we usually do not take this strategy but instead
make the theory back to the manifestly supersymmetric sub-space,
by giving explicitly (or implicitly) counterterms which 
precisely cancel the appearance of non-supersymmetric terms.
This can be done, at least perturbatively, 
unless the symmetry is anomalous.
}.
The physics remains to be unchanged
as the observables do not depend on the renormalization scheme.

In the conventional RG analysis, where we keep 
the original contents of the fields, the difference 
in the renormalization scheme means a tiny tuning 
of the paths of the (almost) fixed IR and UV points.
For example, any scheme in QCD, sooner or later, 
eventually leads to the divergence of the gauge coupling
and its effective Lagrangian becomes hard to analyze.
However, the extended renormalization scheme, 
allowing the new field contents, provides us
a wider choice of the effective actions.
It is possible to have very different IR limits
which share the exactly same physics.
We already know some examples of such an equivalence
as ``duality'' \cite{Montonen:1977sn}.
It is an interesting question to ask if such a duality can be
viewed as an example of the extended renormalization scheme.

In the following sections, we first perform the conventional
RG transformation of two-dimensional QCD (QCD$_2$). 
Note that the model we take has the continuum limit,
and the physical observables can be directly 
expressed by the bare parameters $m$ and $g$.
Since the non-perturbative solutions with $m$ and $g$
are already known, there is no need to perform RG transformations, 
other than comparing with XQCD.

Then, we introduce XQCD at a finite cut-off scale $\Lambda_{\rm cut}$,
and compare its RG flow with QCD, below that scale.
As will be shown below, our computation uses a lot of approximations and assumptions.
It is only at the one-loop level, 
employing a naive soft cut-off, 
assuming the convergence of the computation  even in the Lorentzian space time,
using truncations of the higher order Lagrangians, and so on.
Nevertheless, we find that the RG flow of this simple model
is theoretically non-trivial and interesting.

\section{RG flow of $\text{QCD}_2$ in the large $N_c$ limit}
\label{sec:QCD2calc}
In this section, we analyze the RG flow of the 't Hooft model or QCD$_2$ itself,
without introducing any auxiliary fields.
As mentioned in the previous section, 
this theory is solvable with the bare Lagrangian
in a well-defined continuum limit, and there is no practical needs to renormalize it.
However, its RG analysis turns out to be quite instructive.
Because of the small number of Feynman diagrams in the large $N_c$ limit,
we find that the counterterm which recovers the Parity symmetry, also recovers
the gauge symmetry of the theory along the RG flow.
Moreover, we find a non-perturbative ``solution'' 
(in a truncated theory space), which reasonably interpolates the
theory in the continuum limit and that at the constituent quark mass.
To our knowledge, such a non-perturbative analysis of RG flow in QCD$_2$ is not known before.



\subsection{One-loop analysis and symmetry}
\label{subsec:1loop}

Our goal is to integrate out the high energy modes of
the quark and gluon fields in QCD$_2$ and
obtain an effective action $S_\Lambda$ at a finite cut-off $\Lambda$.
If we could employ a gauge-invariant regularization,
we expect that $S_\Lambda$ has a similar form 
to the bare action :
\begin{align}
\label{eq:reQCD2action}
S_{\Lambda}
&=
\int d^2x~\left[
-\frac{1}{2}\text{Tr}~(\mathbf{A}_+)_R \partial_-^2 
(\mathbf{A}_+)_R
+\bar{\psi}_R(i\Slash{\partial}-m_R(\Lambda))
\psi_R
-\frac{g_R(\Lambda)}{\sqrt{N_c}}\bar{\psi}_R\mathbf{A}_+\gamma^+\psi_R
+ \cdots\right],
\end{align}
where $(\mathbf{A}_+)_R$ and $\psi_R$ denote the renormalized fields, and
$m_R(\Lambda)$ and $g_R(\Lambda)$ are the renormalized mass and coupling constant.
If the effective action has this form, 
one can re-insert the gauge degrees of freedom to the partition function and
recover a manifestly gauge invariant form of the effective theory.
Here we assume that our regularization smoothly cut off the high energy physics.
In this work, we truncate the higher order terms 
and neglect irrelevant contributions at $O(1/\Lambda^4)$.

\begin{figure}[tbhp]
\begin{center}
\includegraphics[width=3cm]{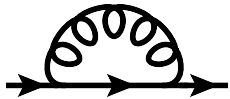}
\caption{The one-loop correction to the quark propagator.}
\label{fig:oneloopSE}
\end{center}
\end{figure}

Since it is difficult to introduce the cut-off in a gauge covariant way,
we usually lose the gauge invariance along the RG flow
even in the truncated theory space.
However, in QCD$_2$ in the light-cone gauge, thanks to the large $N_c$ limit,
the only the one term : $\bar{\psi}\partial_+\gamma^+ \psi$ in the quark kinetic term
obtains quantum corrections (see Fig.~\ref{fig:oneloopSE}).
Because of our choice of the light-cone gauge, this term breaks the Parity symmetry, 
and consequently breaks the gauge symmetry.
By simply adding a counterterm 
or equivalently making a field transformation as we will see below,
one can recover the Parity invariance of the theory,
and the gauge symmetry as well.


Let us demonstrate at the one-loop level how to obtain the effective action $S_\Lambda$ from
our bare action at $\Lambda=\infty$.
It is obtained by expanding the weight $\exp(iS_{\Lambda=\infty})$
in the interaction terms, performing the higher
momentum part above $\Lambda$ of the loop integrals in advance, 
and re-exponentiating them to redefine the new action. 
In our case in the large $N_c$ limit, we have only one term 
non-trivial in this high-mode integration and
we obtain the one-loop result (in momentum space) as
\begin{eqnarray}
\Delta S_{\Lambda}(\mathbf{A}_+, \psi, \bar{\psi}) 
= \int d^2p \left[ -\bar{\psi}\gamma^+ \Delta \Sigma_{\Lambda}(p) \psi \right],
\end{eqnarray}
where 
\begin{equation}
\label{eq:1LSE}
\Delta \Sigma_{\Lambda}(p)
=
4g^2
\int \frac{d^2k}{(2\pi)^2}\left(1-\frac{1}{R_A(-k^2/\Lambda^2)}\right)
\frac{1}{(k_-)^2}
\frac{(p_--k_-)}{(p-k)^2-m^2+i\epsilon}.
\end{equation} 
Here, $R_A(-k^2/\Lambda^2)$ is a smooth function satisfying
the boundary conditions 
\begin{align}
\lim_{k^2\to \infty}\frac{1}{R_A(-k^2/\Lambda^2)} =0,\;\;\;
\lim_{k^2\to 0}\frac{1}{R_A(-k^2/\Lambda^2)} =1.
\end{align}
Note here that we have not renormalized the theory, yet and 
the fields and coupling constants remain to be their bare values
in the continuum limit.

From the Lorentz symmetry, this correction to the quark self-energy
can be decomposed into two parts:
\begin{equation}
\Delta \Sigma_{\Lambda}(p) = -p_+ A(p^2,\Lambda) + B(p^2,\Lambda)/p_-,
\end{equation}
where $A$ and $B$ are regular functions in $p^2$.
The effective action is then
\begin{align}
S_{\Lambda}
=&
\int d^2p~\left[
\frac{1}{2}\text{Tr}~\mathbf{A}_+ p_-^2 
\mathbf{A}_+
-\frac{g}{\sqrt{N_c}}\bar{\psi}\mathbf{A}_+\gamma^+ \psi
\right.\nonumber\\&\left.
+\bar{\psi}\left\{p_-\gamma^-+p_+\gamma^+(1+A(p^2, \Lambda))-(m+\gamma^+ B(p^2,\Lambda)/p_-)\right\}
\psi
\right],
\end{align}
whose Parity symmetry is apparently lost. 
Moreover, one would have a concern about the IR behavior of the term $B(p^2, \Lambda)/p_-$.

However, we can remove these peculiar features by a simple field redefinition :
defining $Z_\psi(p^2,\Lambda)=1/(1+A(p^2, \Lambda))$,
\begin{align}
\label{eq:repsi}
\psi
\equiv
\bigg(
1-\frac{\delta m(p^2, \Lambda)}{2p_-}\gamma^+
\bigg)
Z_\psi(p^2,\Lambda)^{-\frac{\gamma^+\gamma^-}{4}}
\psi_t,
\end{align}
where $\delta m(p^2, \Lambda)$ is the greater solution
of the equation
\begin{align}
2B(p^2,\Lambda)
= 
2\delta m(p^2,\Lambda)m+\delta m(p^2,\Lambda)^2.
\end{align}
With this transformed field $\psi_t$, we obtain a desired form of the effective action,
\begin{align}
S_{\Lambda}
=&
\int d^2p~\left[
\frac{1}{2}\text{Tr}~\mathbf{A}_+ p_-^2 
\mathbf{A}_+
-\frac{g}{\sqrt{N_c}}\bar{\psi_t}\mathbf{A}_+\gamma^+ \psi_t
\right.\nonumber\\&\left.
+
\frac{1}{Z_\psi(p^2,\Lambda)}\bar{\psi_t}\left\{ p_- \gamma^- + p_+ \gamma^+ - \sqrt{Z_\psi(p^2,\Lambda)}(m+\delta m(p^2,\Lambda))\right\} 
\psi_t
\right],
\end{align}
which has both of the Parity and gauge invariances.
We can define the renormalized fields and couplings as
\begin{eqnarray}
(\mathbf{A}_+)_R=\mathbf{A}_+, \;\;\; \psi_R = \sqrt{1/Z_\psi(p^2,\Lambda)}\psi_t,\nonumber\\
m_R(\Lambda) = \sqrt{Z_\psi(p^2,\Lambda)}(m+\delta m(p^2,\Lambda)),\;\;\;
g_R(\Lambda) = Z_\psi(p^2,\Lambda)g,
\end{eqnarray}
to obtain the effective action in Eq.~(\ref{eq:reQCD2action}).
It is interesting to note that the apparent infra-red singularity 
$B(p^2,\Lambda)/p_-$ is converted to the additive mass as the IR cut-off $\delta m(p^2,\Lambda)$.
Also, note that the renormalization of the mass is not linear in $Z_\psi(p^2, \Lambda)$.
These two facts indicate that the quantum correction cannot be considered 
as a simple quark's wave function renormalization.

To recover the gauge symmetry, the renormalization factor $Z_\psi(p^2,\Lambda)$ and the additive mass
$\delta m(p^2,\Lambda)$ should not depend on $p^2$.
In the following computation, we will achieve this by expanding these in $p^2$ around the constituent quark mass $M$,
and set a renormalization condition around the point, stating that
the higher order terms in $(p^2-M^2)/\Lambda^2$ are irrelevant in the low energy region.

There is still one subtlety in the 
IR prescription of the gluon and quark fields.
It is known that the light-cone gauge is sensitive not only to
the IR regularization of the theory, but also to 
the UV regularization function $R_A$ 
when it has a soft cut-off.
Namely, the limit $\Lambda\to \infty$ and the functional integration
may not commute, and the results may differ unless one carefully choose
the IR structure of $R_A$. 
Having a mass gap in QCD,
such an IR subtlety caused by massless gluons
should be unphysical and have no effect on the physical observables.
In fact, the previous works \cite{'tHooft:1974hx,col} 
reported that the meson spectrum and other physical observables are insensitive to 
the choice of IR regularizations.
In this work, however, we would like to keep the IR regularization of the
gluon propagator unchanged from Ref.~\cite{col}, 
in order to make the effect of the UV cut-off $\Lambda$ clearer.

For the gluon propagator, following the prescription by Frishman~\cite{Frishman:1975fd}
we define $R_A$ by
\begin{align}
\frac{i}{k_-^2R_A(k^2/\Lambda^2)}
\label{eq:propIRreg}
=
&~
4i\frac{k_+^2}{k^2+i\epsilon}
\frac{1}{k^2-\mu_{\text{IR}}^2+i\epsilon}
+\pi\epsilon(k_+)
\bigg(
\frac{-2k_+}{\mu^2_{\text{IR}}}
\bigg)
\bigg\{
\delta\bigg(k_--\frac{\mu^2_{\text{IR}}}{2k_+}\bigg)
-
\delta(k_-)
\bigg\}
\nonumber\\
 &~
-4i\frac{k_+^2}{k^2+i\epsilon}
\frac{1}{k^2-\Lambda^2+i\epsilon}
-\pi\epsilon(k_+)
\bigg(
\frac{-2k_+}{\Lambda^2}
\bigg)
\bigg\{
\delta\bigg(k_--\frac{\Lambda^2}{2k_+}\bigg)
-
\delta(k_-)
\bigg\}.
\end{align}
Here, the $\mu_\text{IR}\to 0$ limit has to be taken at the very end of the calculation.
Note that we have introduced IR and UV cut-offs in a symmetric way,
which makes our computation always IR finite,
and the limit $\Lambda\to \infty$ and the path-integration commute.
Without the second and fourth terms, 
the above prescription is similar to the conventional
Pauli-Villars regularization of the gluon field.
The second term corresponds to a 
homogeneous solution of equation of motion
in the $\mu_\text{IR}\to 0$ limit.


Because of the above complication of the choice of IR regularizations,
and Parity and gauge invariances, it is not a good idea to
simply follow the standard procedure explicitly computing one by one, 
in particular, when one wants the computation beyond the one-loop.
Since our bare action is well-defined and there are non-perturbative
results in the continuum limit, it is much easier to
start with the desired effective action Eq.~(\ref{eq:reQCD2action})
and compare the physical observables to those in the continuum limit,
to determine the renormalized coupling and mass.
Namely, in the following, we indirectly determine $S_\Lambda$ by matching
the functional integration from zero to infinity in the bare theory,
and that from zero to $\Lambda$ in the effective theory.

Now let us compute the ($\gamma^-$ component of) quark propagator at the one-loop explicitly,
\begin{align}
\label{eq:quarkprop}
S_\Lambda(p) &= \frac{ip_-}{p^2-m_R(\Lambda)^2-p_- \Sigma_\Lambda (p)+i\epsilon},
\end{align}
where
\begin{align}
\label{eq:1LSigma}
\Sigma_\Lambda(p)
=
&~
4g^2_R(\Lambda)\int
\frac{d^2k}{(2\pi)^2}
\frac{1}{k_-^2 R_A(k^2/\Lambda^2)}
\frac{i(p_--k_-)}{(p-k)^2-m^2_R(\Lambda)+i\epsilon}
\nonumber\\
=&~ -\frac{1}{p_-}\left[
\frac{g_R(\Lambda)^2}{\pi}+
\frac{g_R(\Lambda)^2}{\pi}
\bigg(
\frac{p^2}{\Lambda^2}\log\left|\frac{\Lambda^2}{p^2}\right|
+\frac{m_R(\Lambda)^2}{\Lambda^2}\log\bigg|\frac{\Lambda^2}{m_R(\Lambda)^2}\bigg|
\bigg)
\right]+O(1/\Lambda^4).
\end{align}
Since the (non-perturbative) solution at $\Lambda=\infty$ is known~\cite{col},
\begin{align}
S_\infty (p) &= \frac{ip_-}{p^2-M^2},\;\;\; M^2=m^2-\frac{g^2}{\pi},
\end{align}
we can match this denominator with that of Eq.~(\ref{eq:quarkprop})
up to a renormalization factor,
\begin{align}
p^2-m_R(\Lambda)^2-p_- \Sigma_\Lambda (p) = Z_\psi(\Lambda)^2(p^2-M^2) 
\end{align}
from which we can determine the renormalized quantities as
\begin{align}
Z_\psi^2(\Lambda)=&~\frac{1}{\displaystyle
1-\frac{g^2}{\pi \Lambda^2}\left(\log \bigg|\frac{\Lambda^2}{M^2}\bigg|-1\right)},\nonumber\\
g^2_R(\Lambda) =&~ Z_\psi^2(\Lambda)g^2=\frac{g^2}{\displaystyle
1-\frac{g^2}{\pi \Lambda^2}\left(\log \bigg|\frac{\Lambda^2}{M^2}\bigg|-1\right)},\nonumber\\
m^2_R(\Lambda)=&~
m^2\bigg(1+\frac{2g^2_R(\Lambda)}{\pi\Lambda^2}\log \bigg|\frac{\Lambda^2}{M^2}\bigg|\bigg).
\end{align}

\subsection{Non-perturbative analysis}
\label{subsec:NP}
The above analysis can be easily extended to 
the non-perturbative level.
The self-consistent equation for the rainbow diagram
is given by
\begin{align}
Z^2_\psi(p^2-M^2)=p^2-m_R^2+\frac{g_R^2}{Z_\psi^2\pi}+
\frac{g_R^2}{Z_\psi^2\pi\Lambda^2}\bigg(
\frac{p^2}{\Lambda^2}\log\left|\frac{\Lambda^2}{p^2}\right|
+\frac{M^2}{\Lambda^2}\log\bigg|\frac{\Lambda^2}{M^2}\bigg|
\bigg).
\end{align}
Here we have omitted the arguments of the renormalized quantities for simplicity.
We obtain a set of solutions as follows : 
\begin{align}
&~
\label{eq:rWFresult}
Z^2_\psi(\Lambda)
=1
+\frac{\displaystyle
\frac{g^2}{\pi\Lambda^2}\bigg(\log\bigg|\frac{\Lambda^2}{M^2}\bigg|-1\bigg)}
{\displaystyle 
1-\frac{g^2}{\pi\Lambda^2}\log\bigg|\frac{\Lambda^2}{M^2}\bigg|}, \\
&~
\label{eq:rmassresult}
m^2_R(\Lambda)
=m^2\left(1+
\frac{\displaystyle
\frac{2g^2}{\pi\Lambda^2}\log\bigg|\frac{\Lambda^2}{M^2}\bigg|}
{\displaystyle 
1-\frac{g^2}{\pi\Lambda^2}\log\bigg|\frac{\Lambda^2}{M^2}\bigg|}\right), \\
&~
\label{eq:rcouplingresult}
g^2_R(\Lambda)
=
\frac{\displaystyle
Z^2_\psi(\Lambda)g^2}
{\displaystyle 
1-\frac{g^2}{\pi\Lambda^2}\log\bigg|\frac{\Lambda^2}{M^2}\bigg|}. 
\end{align}

We find here that the chiral symmetry : $\lim_{m\to 0} m_R(\Lambda) = 0$ 
is not compatible with a simple relation 
for the coupling constant $g^2_R(\Lambda)=Z^2_\psi(\Lambda)g^2$.
Since we want to keep the effective action chiral symmetric 
until very low energy limit, we have taken the former relation
$m_R(\Lambda)\propto m$ as our renormalization condition.

\begin{figure}[!tbhp]
\centering
\includegraphics[]{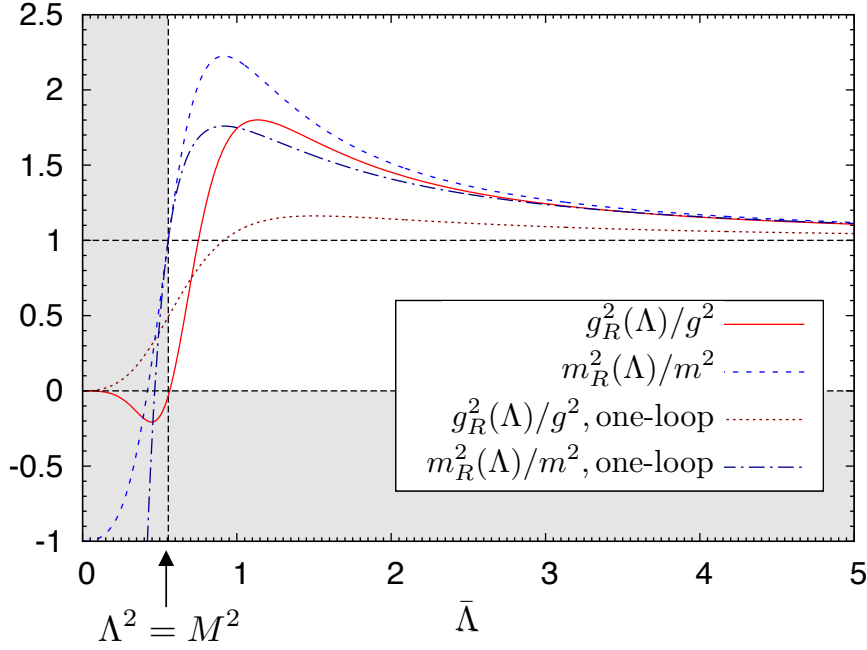}
\caption{The RG running of the mass and coupling of QCD$_2$.
The solid curves are non-perturbative solutions, while the dashed ones are the one-loop results.
The running coupling and mass do not monotonically increase but return to
near the original bare values at $\Lambda\sim M$.
Here, we make all quantities dimensionless using an arbitrarily chosen parameter $\Lambda_0$, and
use $\bar{\Lambda}=\Lambda/\Lambda_0$ for the horizontal axis.
The bare parameters are set to $g/\Lambda_0=1$ and $m/\Lambda_0=0.1$.
}
\label{fig:flow_QCD2}
\end{figure}

The RG running of the mass and coupling constant are given in Fig.~\ref{fig:flow_QCD2}.
Both of the renormalized parameters grow around the starting point as in four-dimensional QCD.
However, it is interesting to note that
they come back to the bare values around the scale of the constituent quark mass,
which is consistent with the fact that their physical
quantities around $\Lambda=M$ should be described by
the bare values $g$ and $m$ again.
For the scale below the constituent quark mass,
there is a region where $g^2_R(\Lambda)$ and $m^2_R(\Lambda)$ go negative.
We do not take this as a serious pathology but just a
failure of our approximation in our crude analysis, including
not taking the threshold effect carefully into account.

\section{RG flow of $\text{XQCD}_2$ in the large $N_c$ limit}
\label{sec:RGXQCD2}

Now let us investigate the RG flow of $\text{XQCD}_2$ in the large $N_c$ limit.
As in the previous section, we truncate our theory 
space to neglect $O(1/\Lambda^4)$ terms.
Also, we require our effective action to be Parity and gauge invariant
(let us just assume that our regularization keeps them by appropriate counterterms).
The large $N_c$ limit also helps to reduce some redundancy of the extended theory space.
For example, the kinetic term of $\mathbf v_\mu$ is never developed.
With this simplification, the most general form of the effective action is 
\begin{align}
S^\text{XQCD}_\Lambda= 
\int d^2p&~\bigg[
\frac{1}{2}\text{Tr}~\mathbf{A}_+ p_-^2\mathbf{A}_+
+\bar{\psi}_R[\Slash{p}-m_R(\Lambda)]\psi_R
-\frac{g_R(\Lambda)}{\sqrt{N_c}}\bar{\psi}_R\mathbf{A}_+\gamma^+ \psi_R \notag \\
&+Z_{\Phi}(\Lambda)\text{Tr}~\Phi^\dag p^2\Phi
-m_\Phi^2(\Lambda)\text{Tr}~\Phi^\dag\Phi
-\frac{\sqrt{2}y(\Lambda)}{\sqrt{N_c}}
\bar{\psi}_R(\Phi P_++\Phi^\dag P_-)\psi_R \notag \\
&
+\frac{1}{2}\lambda^2\text{Tr}~{\mathbf v}_\mu{\mathbf v}^\mu
+i\frac{\alpha\lambda}{\sqrt{N_c}}\frac{Z_\psi(\Lambda)}{Z_\psi(\Lambda_{\rm cut})}
\bar{\psi}_R\Slash{\mathbf{v}}\psi_R
\bigg].
\end{align}
Neglecting the overall normalization of the fields,
our theory space is extended from 2 (with $m_R$ and $g_R$)
to 5 dimensions (since $\alpha$ and $\lambda$ do not run).

As discussed in Sec.~\ref{sec:RGQCD2}, 
we can define a number of new RG schemes in this extended theory space,
by choosing a two-dimensional surface in it.
The simplest (and trivial) scheme is to take the three constraints :
\begin{align}
\label{eq:QCDscheme}
Z_\Phi (\Lambda)=0,\;\;\; m_\Phi^2(\Lambda)=\lambda^2,\;\;\; y(\Lambda)=\alpha \lambda,\;\;\;(\mbox{at any $\Lambda$}),
\end{align}
along the RG flow.
Note that three directions of five-dimensional space are fixed, and thus the RG flow is essentially two-dimensional.
With this scheme, one can always integrate $\Phi$ and $\mathbf v_\mu$ out
and go back to original QCD$_2$ at any scale $\Lambda$.
Since this scheme is exactly equivalent to the scheme in QCD$_2$,
let us call it the ``QCD scheme''.

We are interested in more non-trivial schemes, where the hadronic degrees of freedom
become relevant (let us denote it the ``hadronization scheme'').
Let us require
the same form of the constraints as Eq.~(\ref{eq:QCDscheme})
but only at a point $\Lambda=\Lambda_{\rm cut}$:
\begin{align}
\label{eq:hadronizationscheme}
Z_\Phi (\Lambda_{\rm cut})=0,\;\;\; m_\Phi^2(\Lambda_{\rm cut})=\lambda^2,\;\;\; y(\Lambda_{\rm cut})=\alpha \lambda.
\end{align}
Then, the RG flows can go inside the bulk of the extended five-dimensional space.
Notice that the space of our new RG flow 
still forms a two-dimensional surface,
since it is forced to start 
from the two-dimensional surface at $\Lambda=\Lambda_{\rm cut}$,
and the RG equation is deterministic.
In the following, we compute the RG flow of XQCD in 
this hadronization scheme and compare it with the QCD scheme.

\subsection{One-loop analysis}
\label{subsec:1loopXQCD}

\begin{figure}[htbp]
\begin{center}
\includegraphics[width=5cm]{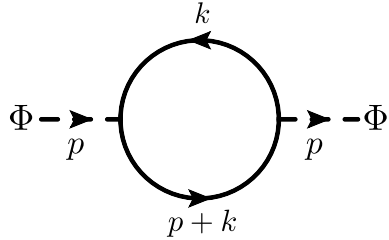}
\caption{$\Phi$'s self energy.} 
\label{fig:qloop}
\end{center}
\end{figure}

\begin{figure}[htbp]
\begin{center}
\includegraphics[width=9cm]{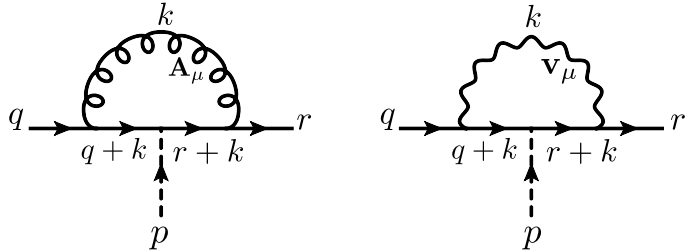}
\caption{Yukawa interaction.} 
\label{fig:yukawa}
\end{center}
\end{figure}

Let us start with the computation at the one-loop.
The three relevant diagrams in the large $N_c$ limit
are the quark self-energy (Fig.~\ref{fig:oneloopSE} the same as QCD$_2$),
the $\Phi$'s self energy (Fig.~\ref{fig:qloop}), 
and Yukawa interaction (Fig.~\ref{fig:yukawa}).

Already at this moment, we can answer to our first question
about the RG flow of the quark and gluon fields in XQCD$_2$.
The three diagrams show that the scalar (and pseudo-scalar) $\Phi$ field
receives quantum corrections from $\psi$ and ${\mathbf v_\mu}$,
but never gives a feedback to them.
Namely, the RG flow of the quark and gluon sector is unchanged.
This result is not what we originally expected :
weakening of the quark and gluon interactions.
It seems that the two-dimension, the light-cone gauge, 
and the large $N_c$ limit simplify the theory too much.
We still expect a non-trivial difference in the case of
four-dimensional QCD with $N_c=3$.

Although there is no essential change in the RG flow of the quark mass 
and gauge coupling, the Feynman diagrams are 
quite different from those in original QCD.
The essential change is in inclusion of the Yukawa interaction,
which makes the mesonic degrees of freedom more relevant,
as will be discussed below.

We begin with computing $\Phi$'s self energy $\Pi(p)$ at the one-loop (Fig.~\ref{fig:qloop}).
When we integrate out high momentum modes between two scales $\Lambda$ and $\Lambda_1$
($\Lambda>\Lambda_1$), we have
\begin{align}
i\Pi(p)
&=
2y^2(\Lambda)\int\frac{d^2k}{(2\pi)^2}
\text{Tr}
\bigg[
P_+\frac{i}{\Slash{k}-m_R(\Lambda)}
P_-\frac{i}{(\Slash{p}+\Slash{k})-m_R(\Lambda)}\bigg] \notag \\
&~~~~~~~~~~~~~~~~~~~~~~~~
\times\bigg[\frac{1}{R_\psi(-k^2/\Lambda^2)}\frac{1}{R_\psi(-(p+k)^2/\Lambda^2)}
-(\Lambda \leftrightarrow \Lambda_1)\bigg]\notag \\
&=\frac{iy^2(\Lambda)}{\pi}\bigg[
\bigg(\frac{1}{\Lambda_1^2}-\frac{1}{\Lambda^2}\bigg)\frac{5}{6}p^2
+\log\bigg(\frac{\Lambda}{\Lambda_1}\bigg)
\bigg]+O(1/\Lambda^4,1/\Lambda^4_1,m^2),
\end{align}
where we have chosen the UV regulator 
\begin{align}
\label{eq:Rpsi}
1/R_\psi (-k^2/\Lambda^2) = \frac{-\Lambda^2}{k^2 - \Lambda^2+i\epsilon}.
\end{align}
These corrections are absorbed in the 
redefinition of $Z_\Phi(\Lambda_1)$ and $m_\Phi(\Lambda_1)$:
\begin{align}
\label{eq:Z_phim_phi}
Z_\Phi(\Lambda_1)
=Z_\Phi(\Lambda)+\frac{5y^2(\Lambda)}{6\pi}\bigg(\frac{1}{\Lambda_1^2}-\frac{1}{\Lambda^2}\bigg),\;\;\;
&m^2_\Phi(\Lambda_1)
=m^2_\Phi(\Lambda)-\frac{y^2(\Lambda)}{\pi}\log\bigg(\frac{\Lambda}{\Lambda_1}\bigg).
\end{align}

Next we turn to the computation of the Yukawa interaction (Fig.~\ref{fig:yukawa}).
The diagram on the left side of Fig.~\ref{fig:yukawa} is 
\begin{align}
&\frac{\sqrt{2}y(\Lambda)g^2_R(\Lambda)}{\sqrt{N_c}}
\int\frac{d^2k}{(2\pi)^2}
\frac{1}{(k_-)^2}\frac{2(q_-+k_-)m_R(\Lambda)}{(q+k)^2-m^2_R(\Lambda)}\frac{1}{(r+k)^2-m^2_R(\Lambda)} \notag \\
&\times\bigg[\frac{1}{R_A(-k^2/\Lambda^2)}\frac{1}{R_\psi(-(q+k)^2/\Lambda^2)}\frac{1}{R_\psi(-(r+k)^2/\Lambda^2)}
-(\Lambda \leftrightarrow \Lambda_1)\bigg]\gamma^+.
\end{align}
However, we neglect this contribution since it is of order $O(1/\Lambda^4)$.

The diagram on the right side of Fig.~\ref{fig:yukawa} is 
\begin{align}
&\frac{\sqrt{2}y(\Lambda)\alpha^2_R(\Lambda)}{\sqrt{N_c}}
\int\frac{d^2k}{(2\pi)^2}
\gamma^\mu\frac{i}{\Slash{q}+\Slash{k}-m_R(\Lambda)}P_\pm\frac{i}{\Slash{r}+\Slash{k}-m_R(\Lambda)}\gamma_\mu \notag \\
&~~~~~~~~~~~~~~~~~~~~~~~~~~
\times\bigg[\frac{1}{R_\psi(-(q+k)^2/\Lambda^2)}\frac{1}{R_\psi(-(r+k)^2/\Lambda^2)}
-(\Lambda \leftrightarrow \Lambda_1)\bigg] \notag \\
=&i\frac{\sqrt{2}y(\Lambda)}{\sqrt{N_c}}
\frac{\alpha_R^2(\Lambda)}{\pi}\log\left(\frac{\Lambda}{\Lambda_1}\right)P_\pm
+O(1/\Lambda^2,1/\Lambda^2_1),
\end{align}
where $P_\pm$ is $P_+$ or $P_-$ when the dashed external line corresponds to $\Phi$ or $\Phi^\dag$ respectively.
$y(\Lambda_1)$ is defined by 
\begin{align}
\label{eq:yukawa_phi}
y(\Lambda_1)
=\frac{Z_{\psi}(\Lambda_1)}{Z_{\psi}(\Lambda)}
\bigg[y(\Lambda)-\frac{y(\Lambda)\alpha_R^2(\Lambda)}{\pi}\log\left(\frac{\Lambda}{\Lambda_1}\right)\bigg],
\end{align}
Here, we have defined $\alpha_R(\Lambda)=\frac{Z_\psi(\Lambda)}{Z_\psi(\Lambda_{\rm cut})}\alpha$.


We obtain the differential RG equations 
by setting $\Lambda_1=\Lambda-d\Lambda$ in Eq.~\eqref{eq:Z_phim_phi} and \eqref{eq:yukawa_phi},
\begin{align}
\frac{dZ_{\Phi}(\Lambda)}{d\Lambda}
=&~
\frac{5y^2(\Lambda)}{6\pi}\frac{d}{d\Lambda}\left(\frac{1}{\Lambda^2}\right), \notag \\
\frac{dm^2_\Phi(\Lambda)}{d\Lambda}
=&~
\frac{y^2(\Lambda)}{\pi}\frac{d}{d\Lambda}\left[\log(\Lambda)\right], \notag \\
\label{eq:RGeqy}
\frac{dy(\Lambda)}{d\Lambda}
=&~
\frac{y(\Lambda)\alpha^2_R(\Lambda)}{\pi}\frac{d}{d\Lambda}\left[\log(\Lambda)\right]
+
\frac{y(\Lambda)g^2_R(\Lambda)}{2\pi}\frac{\partial}{\partial\Lambda}
\left\{\frac{1}{\Lambda^2}
\bigg[\log\left(\frac{\Lambda^2}{M^2}\right)-1\bigg]\right\}.
\end{align}
At the lowest order of perturbation,
the solutions of Eq.~\eqref{eq:RGeqy} are given by
\begin{align}
Z_{\Phi}(\Lambda)
=&~
\frac{5y^2(\Lambda)}{6\pi}\left(\frac{1}{\Lambda^2}-\frac{1}{\Lambda_{\text{cut}}^2}\right)
+O(\Lambda^{-4}), \\
\label{eq:RGsollam}
m^2_\Phi(\Lambda)
=&~
\lambda^2-\frac{y^2(\Lambda)}{\pi}\log\left(\frac{\Lambda_{\text{cut}}}{\Lambda}\right)
+O(\Lambda^{-2}), \\
\label{eq:RGsoly}
y(\Lambda)
=&~
\frac{\alpha\lambda}
{1+\frac{\alpha_R^2(\Lambda)}{\pi}\log\left(\frac{\Lambda_{\text{cut}}}{\Lambda}\right)}
+O(\Lambda^{-2}),
\end{align}
where we have used the initial conditions Eq.~\eqref{eq:hadronizationscheme}.

\begin{figure}[tbhp]
\centering\includegraphics[]{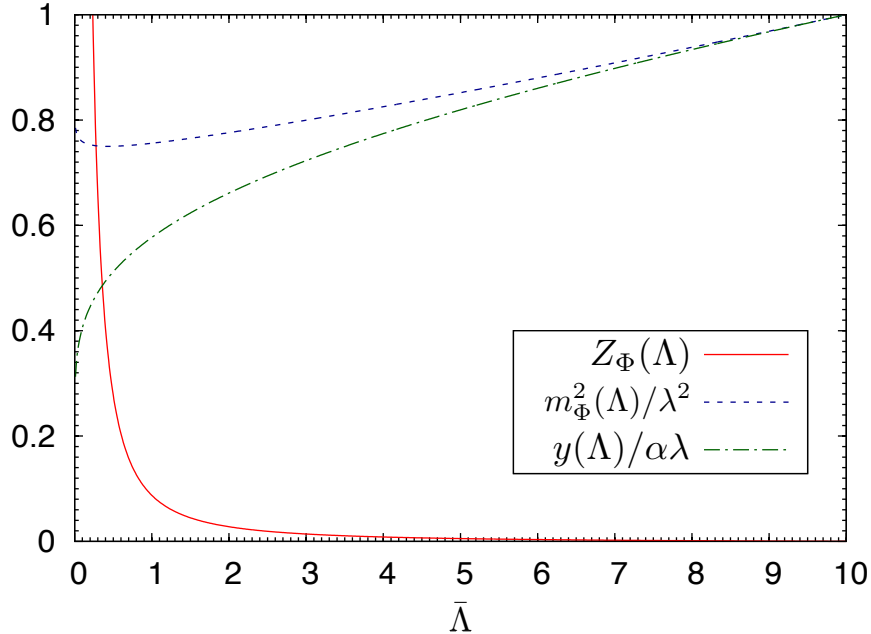}
\caption{RG running of the parameters of XQCD$_2$.
In the same way as Fig.~\ref{fig:flow_QCD2},
all parameters are made dimensionless with the combination of  
the scale parameter $\Lambda_0$.
Here we set $\Lambda_{\text{cut}}/\Lambda_0=10$, $\alpha=1$ 
and $\lambda/\Lambda_0=1$ as their initial conditions.}
\label{fig:flow_XQCD2}
\end{figure}

As is expected, the $\Phi$ field becomes a dynamical
variable, developing its kinetic term, as shown in Fig.~\ref{fig:flow_XQCD2}.

\subsection{Non-perturbative analysis}

Since we have essentially only three types of planar diagrams,
our computation of the RG flow can be, in principle,
 extended to a non-perturbative level.
In particular, as sharing the same quantum numbers as pions,
we expect $\Phi$ to develop a massless pole
in the pseudo-scalar channel.

Unfortunately, we find it not easy to
confirm these expected features  
by simple loop computations even in the large $N_c$ limit.
In fact, this is a well-known problem of the light-cone gauge,
which does not allow any gluonic correction to 
the scalar and pseudo-scalar vertices.
Because of this simple structure, 
the chiral condensate is zero to all order
of loop expansions in the light-cone gauge.
However, the condensate in the 't Hooft model 
is known to be non-zero in the axial gauge 
\cite{Glozman:2012ev}, 
which is inconsistent with its gauge invariance.
Although there have been several proposals 
\cite{Glozman:2012ev,Zhitnitsky:1985um,Chibisov:1995zw} 
to give non-zero contribution from gluons
to the scalar and pseudo-scalar vertices, 
the inconsistency is not yet solved completely as far as we know.

Here we do not go deep inside this controversial issue, 
but simply assume a non-zero expectation value of 
the chiral condensate in the $m\to 0$ limit \cite{Zhitnitsky:1985um}
(we simply change our gauge to the axial gauge 
and come back to the light-cone gauge, assuming the full gauge invariance): 
\begin{equation}
\braket{\bar{\psi}\psi} = -N_c\sqrt{\frac{g^2}{12\pi}}.
\end{equation}
With this assumption, $\Phi$ has also a non-zero VEV (at $m\rightarrow0$) given by 
\begin{equation}
\braket{\Phi}
=
-
\frac{1}{\sqrt{2N_c}}\frac{\alpha}{\lambda}\braket{\bar{\psi}{\psi}}
= \sqrt{\frac{N_c}{24\pi}}\frac{\alpha g}{\lambda}.
\end{equation}
Thus we may re-parametrize $\Phi$ as
\begin{align}
\Phi 
=
\braket{\Phi}
e^{\frac{\sigma+i\pi}{\sqrt{2}}},
\end{align}
where $\sigma$ and $\pi$ are $N_f\times N_f$ hermitian matrices.
With this parametrization, we have
\begin{align}
&~\text{Tr}~\Phi^\dag\Phi
=
\braket{\Phi}^2\text{Tr}~(1+\sqrt{2}\sigma+\sigma^2+\dots), \\
&~\text{Tr}~(\Phi+\Phi^\dag)
=
\braket{\Phi}\text{Tr}(1+\sqrt{2}\sigma+\sigma^2/2-\pi^2/2+\dots).
\end{align}
The linear terms in $\sigma$ in the above two contributions cancel out  in
the Lagrangian.
Combining these equations with Eq.~\eqref{eq:massconvert},
the masses of $\sigma$ and $\pi$ are obtained by
\begin{align}
m^2_\sigma
=
\frac{\alpha^2}{N_c}\braket{\bar{\psi}\psi}^2+O(m)~,~~
m^2_{\pi}
=
\tfrac{1}{2}m\braket{\bar{\psi}\psi}+O(m^2).
\end{align}
Since the mass of $\pi$ is proportional to the quark mass, 
it vanishes in the chiral limit $m \rightarrow0$.
This GMOR relation~\cite{GellMann:1968rz} is kept along the renormalization flow as long as 
our renormalization scheme preserves the chiral symmetry.
For $\sigma$, its mass is proportional to $\Lambda^2$ 
since the mass $Z_\Phi^{-1}(\Lambda)m^2_\Phi(\Lambda)$ is proportional to $\Lambda^2$.
For the quarks, its mass is proportional to $\Lambda$ 
since the Yukawa coupling $Z_\Phi^{-1/2}(\Lambda)y(\Lambda)$ is proportional to $\Lambda$.
(see Subsec.~\ref{subsec:what} (2).)
As we continue to integrate out  high momentum modes,
$\sigma$ and quarks would decouple from the low energy dynamics at some scale,
while $\pi$ continues to contribute to the low energy dynamics.
Eventually the theory is expected to go to the chiral effective theory 
described by the $\pi$ field only and this confirms the low energy hadronic picture. 
We never reach this picture from the RG flow without auxiliary fields.
In this way, the extension of the RG scheme introducing auxiliary fields
gives a different aspect of the theory.

\subsection{What is interesting in the extended RG flow ?}
\label{subsec:what}
Here, we list interesting features and possible applications
of the extended RG flow.
\begin{enumerate}
\item {\it Asymmetry in the RG flow of auxiliary fields}\\
Along the RG flow, we have seen that
$\mathbf{v}_\mu$ remains to be an auxiliary field since
it receives no quantum correction in the large $N_c$ limit.
On the other hand, $\Phi$ acquires its kinetic term 
and becomes dynamical at the low energy.
Clearly the RG flow of $\Phi$ and $\mathbf{v}_\mu$ is asymmetric.
Since we can choose $N_c$ and $N_f$ differently, 
such an asymmetric RG flow is not special for the $N_c=\infty$ limit
but should be common in more general theories.
This is not surprising since there is no symmetry 
between the two fields $\Phi$ and $\mathbf{v}_\mu$.

Here, it is interesting to note that 
the cancellation of $\Phi$ and $\mathbf{v}_\mu$ auxiliary fields
is manifest only at $\Lambda=\Lambda_{\rm cut}$.
If one only had the effective action at $\Lambda \ll \Lambda_{\rm cut}$,
it would be extremely difficult to 
identify that these $\Phi$ and $\mathbf{v}_\mu$ 
originally come from auxiliary fields.
Equivalently, it would be difficult to see that 
this low energy limit of XQCD$_2$ is
equivalent to QCD$_2$, unless one analyzes the high energy behavior
around $\Lambda_{\rm cut}$.

\item {\it Fake UV divergence of auxiliary fields.} \\
In our analysis of the extended RG flow, we have not renormalized 
$\Phi$ so that its coefficient of the kinetic term to be different from unity.
Here let us try the conventional canonical (re)normalization
defining  the renormalized field $\Phi_c$ by 
\begin{align}
\Phi_c \equiv \sqrt{Z_\Phi(\Lambda)}\Phi.
\end{align}
In terms of $\Phi_c$,
its effective mass and effective Yukawa coupling are 
$m_c(\Lambda)\equiv Z^{-1}_\Phi(\Lambda)m^2_\Phi(\Lambda)$ 
and $y_c(\Lambda)\equiv Z^{-\frac{1}{2}}_\Phi(\Lambda)y(\Lambda)$, respectively.
In this normalization, as the renormalized scale $\Lambda$ is
approaching $\Lambda_{\rm cut}$, both of the
mass and Yukawa coupling diverge, since $Z_\Phi(\Lambda_{\rm cut})=0$.

Even for $\Lambda$ much smaller than 
$\Lambda_{\rm cut}$,
the effective mass and the Yukawa coupling behave as
\begin{align}
&Z^{-1}_\Phi(\Lambda)m^2_\Phi(\Lambda)
\sim\frac{6\pi\Lambda^2}{5y^2(\Lambda)}
\bigg[\lambda^2-\frac{y^2(\Lambda)}{\pi}\log\bigg(\frac{\Lambda_{\rm cut}}{\Lambda}\bigg)\bigg], \notag \\
\label{eq:behavior}
&Z^{-\frac{1}{2}}_\Phi(\Lambda)y(\Lambda)\sim\sqrt{\frac{6\pi}{5}}\Lambda,
\end{align}
which look still diverging:
the mass diverges quadratically 
and the coupling diverges linearly
when we go back the RG flow to high energy $\Lambda$.

We, of course, know that our theory is a super-renormalizable
theory and has no divergence.
The appearance of the fake divergence is simply due to
the canonical normalization of the auxiliary degrees of freedom,
and giving an infinite mass to $\Phi$ is consistent
with the fact that the field $\Phi$ becomes a auxiliary field 
and decoupled from the theory.

However, suppose again one only knew the effective
action at low energy $\Lambda \ll \Lambda_{\rm cut}$.
Then, one would find that this theory is very fine-tuned
so that the UV divergence is precisely cancelled at 
$\Lambda=\Lambda_{\rm cut}$ with another 
field $\mathbf{v}_\mu$. 
Since $\mathbf{v}_\mu$ share no symmetry with $\Phi$,
and $\Lambda_{\rm cut}$ has no relation to 
the scale of the original theory, one could think of
the cancellation as very ``unnatural.''


\item {\it  Uniqueness of the ``theory''}\\
One of the essential point of XQCD is that 
the introduction of auxiliary fields 
keeping the partition function unchanged.
The key to achieve such a formulation is 
the Fierz identity (Eq.~\eqref{eq:fierzeu}),
which allows two or three types of auxiliary fields cancelling each other.
However, as discussed in the previous subsection,
we have obtained quite different low energy effective actions 
by considering the RG flow of QCD and XQCD.
In other words, we have two different descriptions 
for the same low energy theory.  

Let us consider more radical set-ups.
There exist infinitely large number of Fierz identities~\cite{Liao:2012uj}.
Moreover, the number of auxiliary fields and
the scale $\Lambda_{\rm cut}$ are arbitrary.
Namely, we have infinite number of the ``extended'' theories
to describe one theory.
Equivalently, we can say that the definition of one theory is not unique.

If there are infinitely many ways or path integrals 
to describe physics, why do we pick up one theory 
as the ``standard'' model ?
Suppose a certain value of $\Lambda_{\rm cut}$, and 
a certain number and kind of introduced auxiliary fields
happened to make all the introduced auxiliary fields weakly coupled and
precise computation of the observables quite easy.
Then one would misidentify the formulation as a ``unique'' theory
and discard other possible descriptions, unless
one finds, by a lucky coincidence, a special re-formulation such as dualities.
It is important to note that there is no physical meaning
on $\Lambda_{\rm cut}$, nor number and kind of auxiliary fields.
The fact that we can introduce these unphysical scale(s),
unphysical flavors, might give some hints for
the long-standing problems in particle physics, 
like the hierarchy problem and problem of three generations.

 \item {\it UV completion for higher spin fields ?}\\
While $\mathbf{v}_\mu$ remains to be an auxiliary field 
in the large $N_c$ limit,
its kinetic term would appear in four-dimensional QCD with $N_c=3$.
In general, the UV completion of a massive vector field is not trivial.
But in XQCD, the UV completion is quite obvious because
the vector field reduces to the auxiliary field at $\Lambda_{\rm cut}$.
The extended RG flow naturally supply the UV completion of the massive vector fields.
Since there are the Fierz identities whose corresponding fields contain higher spin fields~\cite{Liao:2012uj},
the extended RG flow might supply the UV completion of not only massive vector fields but also higher spin fields.
\end{enumerate}

\section{Summary}
\label{sec:conclusion}
In this work,
we have studied the RG flow of QCD$_2$ in the large $N_c$ limit (the 't Hooft model)
and its extension to XQCD$_2$.

For QCD$_2$,
we have found the non-perturbative ``solution" ,
which preserves the Parity symmetry and the gauge symmetry along the RG flow.
As seen in Fig.~\ref{fig:flow_QCD2},
the values of the effective mass and coupling 
grow around the starting point and then return to the bare values
around the scale of the constituent quark mass.
We can see that the RG flow of QCD$_2$ smoothly interpolates 
the theory in the continuum limit and that at the constituent quark mass.

For XQCD$_2$,
although our specific analysis of the RG flow is at the one-loop level,
we have found non-trivial and interesting pictures of 
the RG flow with auxiliary fields.
By the introduction of the auxiliary fields,
the parameter space of the theory is extended from the original one.
However, as the auxiliary fields should not change the physics,
the RG flow in the extended parameter space forms a surface
whose dimension is the same as the original parameter space. 
The choice of the surface is not unique and
corresponds to the choice of the (extended) renormalization scheme.

In Sec.~\ref{sec:RGXQCD2},
we have compared two schemes in the RG flow of XQCD.
One is the ``QCD scheme"  where
all auxiliary fields remain to be non-dynamical
and equivalent to the RG flow of original QCD.
They can be removed at any scale of $\Lambda$ from the theory and 
we simply go back to the original QCD effective action.

Another is  the ``hadronization scheme", 
where the scalar auxiliary field $\Phi$ becomes dynamical 
while the vector auxiliary field $\mathbf{v}_\mu$ 
still remains to be an auxiliary.
Assuming the chiral symmetry breaking in the 't Hooft model,
the constituent quarks and the massive scalar fields 
obtain a mass $\sim g$ or $\Lambda$.
The only pions remain near massless and
 relevant in the low energy region.
This confirms the hadoronic picture of QCD.
Since ``QCD scheme" does not show this picture,
we emphasize that we can never realize such a picture  
without taking into account the RG flow with auxiliary field,
in other words, without adding the new elementary field 
which contains the pion degrees of freedom to QCD.

\section*{Acknowledgments}
We thank Kazuhiko Kamikado, David B. Kaplan, Kengo Kikuchi, Tetsuya Onogi, 
and Masatoshi Yamada for fruitful discussions and useful comments.
We also thank the Yukawa Institute for Theoretical Physics, 
Kyoto University. Discussions during the YITP workshop YITP-T-14-03 
on “Hadrons and Hadron Interactions in QCD" were useful to complete this work.
This work is supported in part by the Grand-in-Aid of the Japanese Ministry of Education 
No.25800147, 26247043 (H.F.), and No. 15J01081 (R.Y.).

\bibliographystyle{ptephy}
\bibliography{bib}

\begin{thebibliography}{10}

\bibitem{Kaplan:2013dca}
David~B. Kaplan (2013),  {{arXiv:1306.5818}}.

\bibitem{Chodos:1975ix}
Alan Chodos and Charles~B. Thorn, Phys.Rev., {\bf D12}, 2733 (1975).

\bibitem{'tHooft:1974hx}
Gerard 't~Hooft, Nucl.Phys., {\bf B75}, 461 (1974).

\bibitem{Nambu:1961tp}
Yoichiro Nambu and G.~Jona-Lasinio, Phys.Rev., {\bf 122}, 345--358 (1961).

\bibitem{Braun:2014ata}
Jens Braun, Leonard Fister, Jan~M. Pawlowski, and Fabian Rennecke (2014),
  {{arXiv:1412.1045}}.

\bibitem{Brower:1994sw}
Richard~C. Brower, Yue Shen, and Chung-I Tan, Nucl.Phys.Proc.Suppl., {\bf 34},
  210--212 (1994),  {{arXiv:hep-lat/9403011}}.

\bibitem{Brower:1995vf}
R.C. Brower, K.~Orginos, and C.I. Tan, Nucl.Phys.Proc.Suppl., {\bf 42}, 42--48
  (1995),  {{arXiv:hep-lat/9501026}}.

\bibitem{Kogut:2004ia}
J.B. Kogut and D.K. Sinclair (2004),  {{arXiv:hep-lat/0408003}}.

\bibitem{GellMann:1964nj}
Murray Gell-Mann, Phys.Lett., {\bf 8}, 214--215 (1964).

\bibitem{Zweig:1981pd}
G.~Zweig and S.~An, CERN Report, {\bf 8419} (1964).

\bibitem{col}
Sidney~R. Coleman, {Cambridge University Press} (1988).

\bibitem{Bars:1977ud}
I.~Bars and Michael~B. Green, Phys.Rev., {\bf D17}, 537 (1978).

\bibitem{Li:1987hx}
M.~Li, L.~Wilets, and M.C. Birse, J.Phys., {\bf G13}, 915--923 (1987).

\bibitem{Montonen:1977sn}
C.~Montonen and David~I. Olive, Phys.Lett., {\bf B72}, 117 (1977).

\bibitem{Frishman:1975fd}
Y.~Frishman, Nucl.Phys., {\bf B148}, 74 (1979).

\bibitem{Glozman:2012ev}
L.~Ya. Glozman, V.K. Sazonov, M.~Shifman, and R.F. Wagenbrunn, Phys.Rev., {\bf
  D85}, 094030 (2012),  {{arXiv:1201.5814}}.

\bibitem{Zhitnitsky:1985um}
A.R. Zhitnitsky, Phys.Lett., {\bf B165}, 405--409 (1985).

\bibitem{Chibisov:1995zw}
Boris Chibisov and Ariel~R. Zhitnitsky, Phys.Lett., {\bf B362}, 105--112
  (1995),  {{arXiv:hep-ph/9502258}}.

\bibitem{GellMann:1968rz}
Murray Gell-Mann, R.J. Oakes, and B.~Renner, Phys.Rev., {\bf 175}, 2195--2199
  (1968).

\bibitem{Liao:2012uj}
Yi~Liao and Ji-Yuan Liu, Eur.Phys.J.Plus, {\bf 127}, 121 (2012),
  {{arXiv:1206.5141}}.

\end{thebibliography}

\end{document}